\documentclass[12pt]{article}
\usepackage[latin1]{inputenc}

 \usepackage{amsmath}
 \usepackage{amsfonts}
 \usepackage{amsthm}
 \usepackage{graphicx}
 \usepackage{xcolor}
\newtheorem{definition}{D\'efinition}
\newtheorem{proposition}{Proposition}
\newtheorem{remark}{Remarque}

\theoremstyle{definition}

\DeclareMathOperator{\Tr}{Tr}

\DeclareMathOperator{\RR}{\mathbb{R}}

\begin{document}


\begin{center}
{\Large
	{\sc  Analyse discriminante matricielle descriptive. Application à l'étude de signaux EEG }
}
\bigskip

 Juliette Spinnato$^{1,3,4}$ \& Marie-Christine Roubaud$^{1}$  \& Margaux Perrin $^{2}$ \& Emmanuel Maby $^{2}$  \& Jeremie Mattout $^{2}$ \& Boris Burle $^{3}$ \& Bruno Torrésani $^{1}$ 
\bigskip

{\it
$^{1}$ Aix-Marseille Université, CNRS, Centrale Marseille, I2M, UMR 7373, 13453 Marseille  \\
$^{2}$ Université Lyon 1,CNRS, INSERM U1028, CRNL, UMR 5292, 69000 Lyon \\
$^{3}$ Aix-Marseille Université, CNRS, LNC, UMR 7291, 13331 Marseille \\
$^{4}$ juliette.spinnato@univ-amu.fr

}
\end{center}
\bigskip


{\bf R\'esum\'e.} 
Nous nous intéressons à l'approche descriptive de l'analyse discriminante
linéaire de données matricielles dans le cas binaire.
Sous l'hypothèse de séparabilité de la variabilité des lignes de celle des colonnes,  les combinaisons linéaires des lignes et des colonnes {\it les plus discriminantes} sont déterminées par la décomposition en valeurs singulières de la différence des moyennes des deux classes en munissant  les espaces des lignes et des colonnes de la  métrique de Mahalanobis. 
Cette approche permet d'obtenir des représentations des données dans des plans factoriels et de dégager des composantes discriminantes.   
Une application à des signaux d'électroencéphalographie multi-capteurs illustre la pertinence de la méthode.

\smallskip

{\bf Mots-cl\'es.} Analyse discriminante linéaire, données matricielles, matrice de covariance séparable,  décomposition en valeurs singulières, signaux EEG  
\bigskip

{\bf Abstract.} 
We focus on the descriptive approach to linear discriminant analysis for matrix-variate data  in the binary case.
Under a separability assumption on row and column variability, the {\it most discriminant} linear combinations of rows and columns are determined by the singular value decomposition of the  difference of the class-averages with the Mahalanobis metric in the row and column spaces. 
This approach provides data representations of data in two-dimensional or three-dimensional plots and singles out discriminant components.
An application to electroencephalographic multi-sensor signals illustrates the relevance of the method.

\smallskip

{\bf Keywords.} Linear discriminant analysis, matrix-variate data, separable covariance matrix, singular value decomposition, EEG signal



\vspace{-5pt}
\section{Introduction}
\vspace{-5pt}
L'analyse discriminante linéaire (LDA) de données structurées de type matriciel est étudiée  dans de nombreux contextes  notamment en classification de signaux multi-capteurs telles que les données électroencéphalographiques (EEG) (e.g. Guitiérrez \& Escalona-Vargas, \-2010). Une approche simple pourrait être de vectoriser les données et d'utiliser la LDA classique sans tenir compte de leur structure. Cependant, 
les vecteurs ainsi construits sont généralement de très grande dimension et la LDA classique est mise à défaut. En effet, 
le nombre de paramètres à estimer est souvent supérieur au nombre d'observations et l'estimation de la matrice de covariance est donc de très mauvaise qualité.  Afin de réduire le nombre de paramètres à estimer il est usuel d'introduire une hypothèse de séparabilité entre la variabilité des lignes et celle des colonnes.  Ceci permet de prendre en compte la structure des données par  un modèle de covariance  simple sous la forme du produit de Kronecker de la matrice de covariance des lignes et de celle des colonnes (e.g. Akdemier $\&$ Gupta,~(2011) et les références citées).  
Cette hypothèse est raisonnable dans plusieurs situations et notamment dans l'étude des signaux EEG, cadre applicatif de ce travail (e.g. Bijma \textit{et al.},~(2005), Friston \textit{et al.},~(2005) et Mahanta \textit{et al.},~(2012)). 

L'objectif de ce travail est de proposer  une méthode de détermination  de  combinaisons linéaires de lignes et de colonnes {\it les plus discriminantes} entre deux classes en tenant compte de la structure des données sous l'hypothèse de séparabilité.  Notre approche repose sur la décomposition en valeurs singulières (SVD) de la différence des moyennes  en munissant  les espaces  des lignes et des colonnes de la  métrique de Mahalanobis. Ceci nous permet d'extraire conjointement  les combinaisons  {\it les plus discriminantes } dans ces deux espaces et d'en déduire des représentations graphiques de ces  composantes  ainsi que des visualisations des données  matricielles dans des sous-espaces de dimension réduite. Une approche similaire a déjà été appliquée  à l'extraction de caractéristiques discriminantes  des potentiels d'erreur dans les signaux EEG  dans un contexte de classification binaire par Spinnato \textit{et al.}~(2014).  Dans ce travail nous développons l'aspect descriptif de la méthode et étudions son apport pour des signaux EEG dans un paradigme d'Interface Cerveau-Machine de type P300 Speller (Farwell \& Donchin,~1988).  
 
La structure de cet article est la suivante. Nous introduirons tout d'abord quelques notations et  définitions. Nous présenterons ensuite l'approche matricielle  descriptive de la LDA binaire sous l'hypothèse de séparabilité et son apport en termes de réduction de dimension et d'extraction de caractéristiques discriminantes.  Enfin nous montrerons la pertinence de la méthode pour  l'étude des caractéristiques discriminantes  de potentiels évoqués en EEG.
   
\vspace{-10pt}  
\paragraph{Notations et définitions.} Notons $\RR^{K\times J}$ l'espace des matrices réelles à $K$ lignes et $J$ colonnes.  
Soient $M\in\RR^{J\times J}$ et $D \in\RR^{K\times K}$ deux matrices symétriques et définies positives. L'espace des lignes    $\RR^J$ est dit muni de la  métrique  $M$ si le  produit scalaire sur cet espace  est défini par $\langle x,y \rangle_M= x^\prime M y$ pour tout $x,y\in\RR^J$. De manière similaire, l'espace des colonnes $\RR^K$ est muni de la métrique $D$.  
 \begin{definition}[Produit scalaire, norme matricielle]
 Soient  $X$ et $Y$ des matrices $\in \RR^{K\times J}$, 
 \vspace{-10pt}
 \begin{equation}
 \label{psmat}
 \langle X,Y \rangle_{M,D} = \Tr(XMY^\prime D)\; \mbox{ et } \; \|X \|^2_{M,D} = \Tr(XMX^\prime D)\,,
 \end{equation}
où $\Tr$ désigne l'opérateur trace et $X^\prime $ la matrice transposée de $X$.
\end{definition} 
Sachant que 
$\Tr(A^\prime B C D)= \mathsf{vec}(A)^\prime(D\otimes B)\mathsf{vec}(C) \,, $ où $\mathsf{vec}(A)$ d\'esigne la vectorisation de la matrice $A$ par concaténation de ses colonnes et $ \otimes $ le produit de Kronecker, on obtient l'équivalence entre le produit scalaire défini en (\ref{psmat}) sur $\RR^{K\times J}$  et celui défini sur l'espace $\RR^{KJ}$ muni de la métrique $D\otimes M$:
\begin{equation}
 \label{psmatvec}
 \langle X,Y \rangle_{M,D} = \mathsf{vec}(X^\prime)^\prime (D\otimes M)  \mathsf{vec}(Y^\prime) = 
 \langle \mathsf{vec}(X^\prime),\mathsf{vec}(Y^\prime) \rangle_{D\otimes M} \,.
 \end{equation}

\vspace{-20pt}
\section{ Analyse discriminante matricielle descriptive sous l'hypothèse de séparabilité}

\subsection*{Mod\`ele et estimation }
\label{sse:modele}

Nous nous int\'eressons à  l'analyse discriminante binaire dans le cas de donn\'ees structur\'ees sous forme matricielle. Dans chacune des deux classes $c=1,2$, les observations $X^i_c\in\RR^{K\times J}$ pour $i=1,\ldots,n$, sont considérées comme des r\'ealisations i.i.d. d'une matrice al\'eatoire $X_c$  de moyenne $\mu_c$ dépendante de la classe. De plus, nous supposons que les  matrices de covariance des lignes $\Sigma_L\in\RR^{K\times K}$ et des colonnes  $\Sigma_R \in\RR^{J\times J}$ sont indépendantes de $c$, inversibles  et telles que 
\vspace{-6pt}
\begin{equation}
\label{eq:hyp_sep}
 \Sigma= \Sigma_L \otimes \Sigma_R  \\,
\end{equation}
où $\Sigma\in\RR^{KJ\times KJ} $ est la matrice de covariance de $\mathsf{vec}(X_c^\prime)\in \mathbb{R}^{KJ}$ correspondant à  la vectorisation de la matrice $X_c^\prime$. \\

Soient $n_1$ et $n_2$ les effectifs des classes 1 et 2. Posons $n=n_1+n_2$. 
Comme définie par  Dutilleul P.~(1999), l'estimation des matrices de covariance $\Sigma_L$ et $\Sigma_R$  s'effectue de manière itérative via une normalisation des $J$ colonnes et des $K$ lignes: 
\vspace{-10pt}
\begin{eqnarray}
 & S_W^L &=  \frac{1}{nJ} \sum_{c=1}^2 \sum_{i=1}^{n_c} (X^i_c - \bar{X}_c)(S_W^R)^{-1}(X^i_c - \bar{X}_c)' \,, \label{eq:SL} \\  
 & S_W^R &= \frac{1}{nK}  \sum_{c=1}^2 \sum_{i=1}^{n_c}(X^i_c - \bar{X}_c)'(S_W^L)^{-1}(X^i_c - \bar{X}_c) \,, \label{eq:SR}
\end{eqnarray}

\noindent o\`u  $\bar{X}_c= \frac{1}{n_c} \sum_{i=1}^{n_c} X^i_c $. Nous en déduisons l'estimation  de la matrice de covariance $\Sigma$ :
\vspace{-6pt}
\begin{equation}
S_W = S_W^L \otimes S_W^R.
\end{equation}

\begin{remark}[Identifiabilit\'e]
Pour tout $\kappa \neq 0$, notons
que $S_W^L \otimes S_W^R = \kappa S_W^L \otimes \kappa^{-1} S_W^L$. Afin de r\'esoudre
ce probl\`eme d'identifiabilit\'e la norme de $S_W^R$ est fixée \`a 1 dans l'estimation.
\end{remark} 

\vspace{-20pt}
\subsection*{Décomposition en valeurs singulières de la différence des moyennes}

Rechercher les combinaisons linéaires des lignes et des colonnes les plus discriminantes  équivaut à  décomposer en valeurs singulières  la différence des moyennes empiriques des deux classes  en munissant  $\RR^J$ l'espace des lignes (resp. $\RR^K$, l'espace des colonnes)  de la métrique $M=(S_W^R)^{-1}$ (resp. $D=(S_W^L)^{-1} $). 

\begin{proposition}
Soit $Q$ le rang de la matrice $\bar{X}_1 - \bar X_2 \in \RR^{K\times J}$.
  \vspace{-10pt}
\begin{equation}
\label{SVD}
   \bar X_1 - \bar X_2  =U\Lambda^{1\over 2}V^\prime= \sum_{q=1}^Q \sqrt{\lambda_q} u_qv_q^\prime \,, \quad \hbox{où}
   \vspace{-10pt}
 \end{equation}
   \vspace{-10pt}
 \item[-] $U \in\RR^{K\times Q}$ a pour colonnes les vecteurs propres $D$-orthonormés  de la matrice $D$-symétrique semi-définie positive $(\bar{X_1} - \bar{X_2})M(\bar{X_1} - \bar{X_2})^\prime D$ associés aux $Q$ valeurs propres non nulles $\lambda_q$ rangées dans un ordre décroissant dans la matrice diagonale $\Lambda$\,.
\item[-] $V \in\RR^{J\times Q}$ a pour colonnes les vecteurs propres $M$-orthonormés  de la matrice $M$-symétrique semi-définie positive $(\bar{X_1} - \bar{X_2})^\prime D(\bar{X_1} - \bar{X_2})M $ associés aux valeurs propres non nulles $\lambda_q$  avec \vspace{-10pt}
\begin{equation}
\label{UV}
V= (\bar{X_1} - \bar{X_2})^\prime D U \Lambda^{-\frac{1}{2}}\,.
\end{equation} 
\end{proposition}
\vspace{-5pt}
D'après \eqref{psmatvec},   l'espace $\RR^{KJ}$  est muni de la métrique $D\otimes M= S_W^{-1}$.
A partir de \eqref{SVD}, sachant que $ \mathsf{vec}(u_qv_q^\prime) = u_q\otimes v_q$,   nous obtenons une décomposition  pour  la différence des moyennes vectorisées et pour la distance de Mahalanobis entre ces moyennes\,: 
\begin{proposition} Soient $u_q\otimes v_q$, $q=1,\ldots Q$, des vecteurs  $S_W^{-1}$-orthonormés de  
$\RR^{KJ}$. Alors  \vspace{-10pt} 
\begin{equation}    
\label{svd vec}
 \mathsf{vec}(\bar X_1^\prime -\bar X_2^\prime) =  \sum_{q=1}^Q \sqrt{\lambda_q} (u_q\otimes v_q )\,, \quad \hbox{et}
\end{equation}
\vspace{-15pt}
\begin{equation}\label{declambda}
\| \mathsf{vec}(\bar X_1^\prime - \bar X_2^\prime)\|^2_{S_W^{-1}} =   
\sum_{q=1}^Q \lambda_q \; \mbox{ avec }\; \lambda_q= |\langle \mathsf{vec}( \bar{X_1}^\prime  - \bar{X_2}^\prime),  u_q\otimes v_q \rangle_{S_W^{-1}}|^2\,.
\end{equation}
\end{proposition}
\vspace{-20pt}

\paragraph{Projection dans l'espace lignes-colonnes $\RR^{KJ}$.} 
Les  coordonnées des observations  $X^i$ vectorisées  sur les axes  engendrés par les vecteurs $ u_q\otimes v_q $ pour $q=1,\ldots, Q$ sont données par $  \langle \mathsf{vec}( (X^i)^\prime ),  u_q\otimes v_q \rangle_{S_W^{-1}}$. 
Pour la différence de moyennes, on obtient de plus  l'erreur d'approximation.  
\begin{proposition}
\label{prop:erreurapprox}
Soit $ \mathsf{proj}_{E_r}\left( \mathsf{vec}(\bar X_1^\prime-\bar X_2^\prime)\right) =  \sum_{q=1}^r \sqrt{\lambda_q} (u_q\otimes v_q )$ la projection de la différence des moyennes sur $E_r$ le sous-espace engendré par les $r$ premiers vecteurs $ u_q\otimes v_q $.   

\vspace{-10pt}
\begin{equation}
\label{erreur}
 \| \mathsf{vec}(\bar X_1^\prime - \bar{X_2}^\prime) - \mathsf{proj}_{E_r}\left( \mathsf{vec}(\bar X_1^\prime-\bar X_2^\prime)\right) \|^2 _{S_W^{-1}} =   \lambda_{r+1}+\lambda_{r+2}+ \cdots +\lambda_Q \,. 
\end{equation}
\end{proposition}

\vspace{-15pt}
\paragraph{Projection dans l'espace des lignes ($\RR^J$) et   des colonnes ($\RR^K$).}
Les coordonnées des $K$ lignes (resp. des $J$ colonnes)  de $X^i$ sur les axes engendrés par les vecteurs $v_q$, $q=1,\ldots, Q$ (resp. $u_q$,  $q=1,\ldots, Q$) sont données respectivement par 

\vspace{-10pt}
\begin{equation}
\label{coordlignescol }
 X^i M v_q  = X^i (S_W^R)^{-1}  v_q\; \mbox{ et } \; (X^i)^\prime D u_q = (X^i)^\prime (S_W^L)^{-1}  u_q\,.  
\end{equation}

\vspace{-18pt}
\section{Application: étude des composantes discriminantes}

\vspace{-5pt}
\paragraph{Données et pré-traitement.} La méthode présentée est appliquée à l'analyse en composantes discriminantes de potentiels évoqués corticaux. 
Ces signaux EEG ont été acquis dans le cadre du protocole  P300 Speller~(Farwell \& Donchin,~1988) dont le but est d'épeler des lettres en décodant l'activité cérébrale.

Ce dispositif consiste à afficher une matrice $6 \times 6$  composée de l'alphabet et des chiffres. Le participant souhaitant écrire une lettre doit se concentrer sur la case correspondante dans la matrice, dont les lignes et les colonnes sont successivement et aléatoirement {\it flashées}. Lorsque la ligne ou la colonne contenant la lettre cible est {\it flashée}, plusieurs composantes électrophysiologiques sont générées et en particulier une composante visuelle (l'onde négative N1) associée au stimulus visuel, suivie d'une seconde composante liée à la  détection du stimulus (l'onde positive P300). 
Deux classes sont donc considérées: les cibles ($c=1$) et les non-cibles ($c=2$).
 Afin d'épeler une lettre, chaque ligne et chaque colonne est {\it flashée} $3$ fois, ce qui correspond à un total de 36 $flashs$ dont $6$ sont des cibles. Dans la suite, nous basons nos résultats sur un échantillon de 20 lettres épelées par un participant dans l'expérience menée par Perrin \textit{et al.}~(2011).

Chaque essai $i$ pour $i=1,\ldots n$, avec $n=720$, correspond à une portion du si\-gnal enregistré dans l'intervalle de temps $[-100 ms;+900ms]$ où zéro représente l'instant du flash (fréquence d'échantillonnage: $1000Hz$).
Une transformation en ondelettes discrètes est appliquée en pré-traitement  afin de résumer l'information temporelle. Un filtre de Daubechies D8 est appliqué sur 5 niveaux de décomposition, et seuls les coefficients supérieurs à  leur moyenne ont été conservés pour l'analyse.  
Les données pré-traitées sont donc de la forme  $X_c^{i}  \in \RR^{K\times J}$, où  $K$ est le nombre de coefficients sélectionnés ($K = 28$), $J$ est  le nombre d'électrodes ($J = 32$) et $c$ précise la classe. 

\vspace{-10pt}
\begin{figure}[htbp]
\begin{minipage}[c]{.45\linewidth}
\begin{center}
\includegraphics[width=1\textwidth]{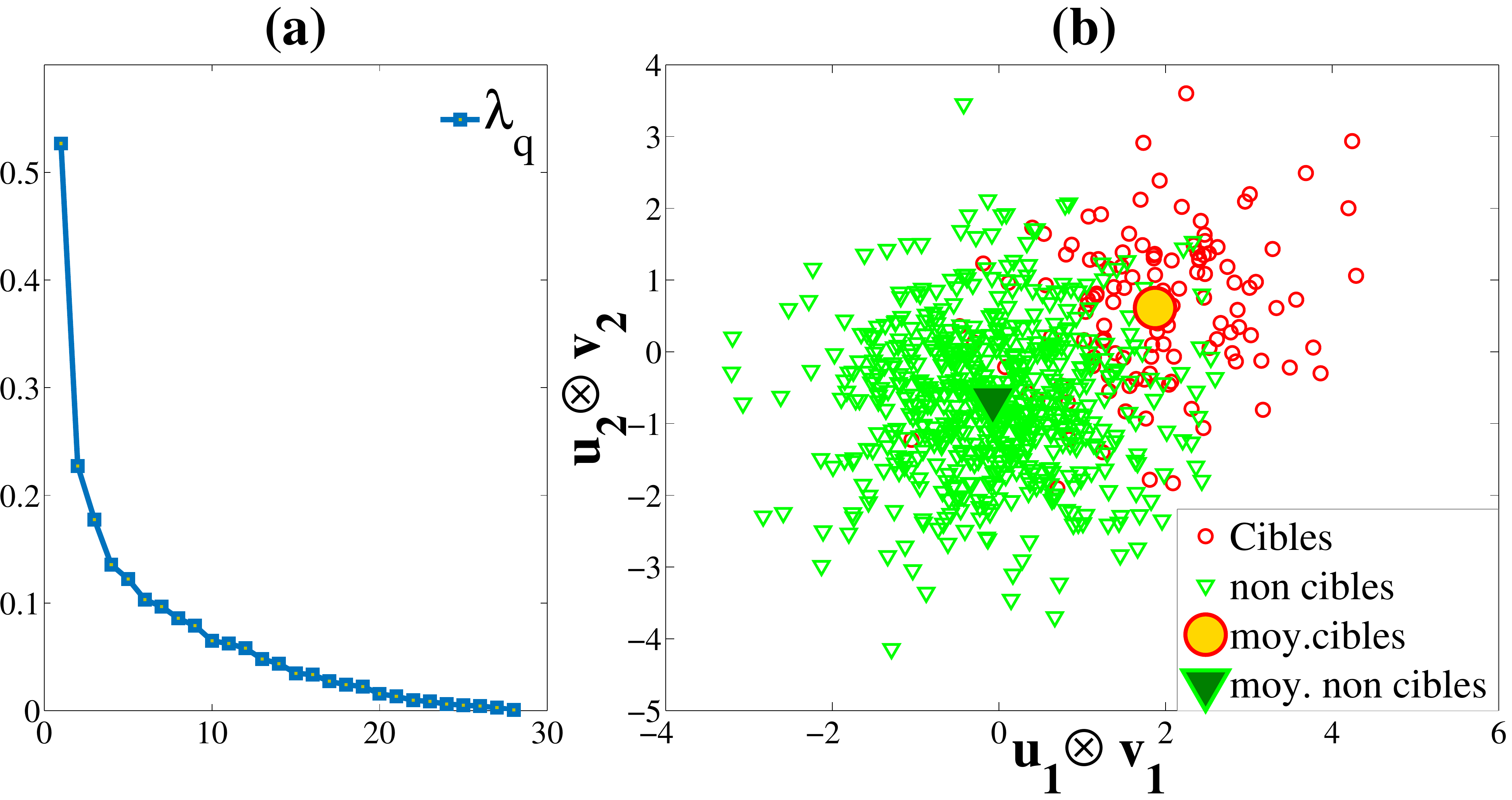} 
\caption{\small Valeurs propres (a) et représentation des observations et des moyennes dans le premier plan spatio-temporel (b).}
\label{fig:repres_ind}
\end{center}
\end{minipage}
\hfill
\begin{minipage}[c]{.45\linewidth}
\begin{center}
\includegraphics[width=1\textwidth]{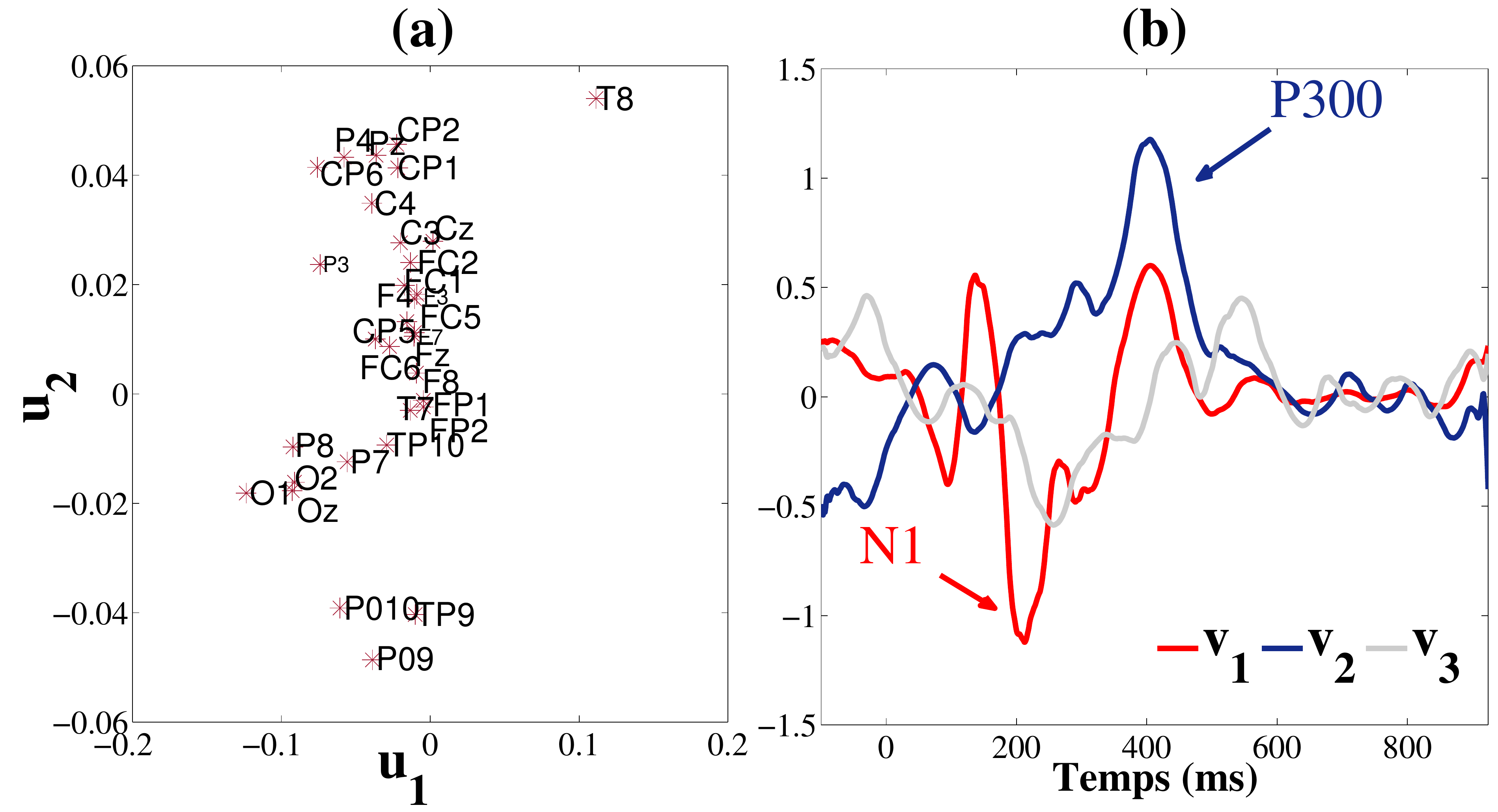} 
\caption{\small Composantes discriminantes spatiales (a) et temporelles (b)  de la différence des moyennes des 2 classes. \hspace{3cm} \ }
\label{fig:variables_discrim}
\end{center}
\end{minipage}
\end{figure}

\vspace{-23pt}
\paragraph{Projection de données dans l'espace spatio-temporel.}
La figure~\ref{fig:repres_ind}(a) représente l'éboulis des $28$ valeurs propres non nulles. 
Dans un objectif de réduction de dimension, l'identification  d'un "coude" suggère de choisir le sous-espace de projection $E_r$ de dimension $r=4$. Cette approximation peut être particulièrement
utile en classification.
La figure~\ref{fig:repres_ind}(b)  représente la projection des observations sur le premier plan factoriel engendré par $u_1\otimes v_1$ et $u_2\otimes v_2$ ainsi que les deux moyennes  
des cibles et des non-cibles. Cette représentation permet de visualiser la séparation des deux classes dans ce sous-espace et de distinguer les observations atypiques. 

\vspace{-15pt}
\paragraph{Analyse des composantes discriminantes.}  
La figure~\ref{fig:variables_discrim}(b) représente les trois pre\-miè\-res composantes temporelles  discriminantes. Elles ont été obtenues par synthèse d'ondelettes à partir des projections des différences des coefficients moyens entre les 
classes cible et non-cible sur les axes engendrés par $v_1$, $v_2$ et $v_3$.  Les deux principales composantes associées respectivement à $v_1$ et $v_2$, deux vecteurs $(S_W^R)^{-1}$-orthogonaux, apparaissent si\-mi\-lai\-res à l'onde N1 et à  l'onde P300. Ceci semble donc indiquer une forme de découplage de ces deux composantes.
La figure~\ref{fig:variables_discrim}(a) représente la projection dans le plan $(u_1,u_2)$ des  différences des moyennes des classes mesurées sur chaque électrode.
A partir de cette représentation des groupes distincts d'électrodes peuvent être associés  à chacune des composantes précédentes. De plus, on peut détecter des comportements atypiques, comme celui de l'électrode T8. Le comportement
singulier de cette électrode située sur la tempe droite est connu (artefacts) et il n'est pas rare de devoir l'exclure des analyses.

\vspace{-10pt}
\section{Conclusion}
\vspace{-10pt}
 
Sous l'hypothèse de séparabilité, la formulation de l'analyse discriminante matricielle binaire comme une SVD de la différence des moyennes des classes offre un cadre générique permettant d'utiliser les propriétés classiques de la SVD.
Notamment, la dualité conduit à une analyse conjointe ligne-colonne qui fournit des descripteurs dans les deux domaines simultanément. D'autre part, la réduction de dimension induite peut être 
utilisée en classification.
Dans ce travail nous avons montré la pertinence de cette méthode pour l'analyse spatio-temporelle des signaux EEG.
La méthode permet  d'extraire et d'analyser distinctement deux composantes (identifiées aux ondes N1 et P300) qui semblent être découplées autant dans le domaine temporel qu'en termes de localisation spatiale.

\vspace{-10pt}
\section*{Bibliographie}
\vspace{-9pt}

\noindent [1] Guitiérrez, D., et Escalona-Vargas, D.I. (2010), {\it EEG data classification through signal spatial
redistribution and optimized linear discriminants}, {Comp.\/Meth.\/Prog.\/Bio.}, 97, 39-47.

\noindent [2] Akdemir, D. et Gupta, A.K. (2011), \textit{Array variate random variables with multiway Kronecker delta covariance matrix structure}, {Journal of Algebric Statistics}, 2(1), 98-113.

\noindent [3] Bijma, F., de Munck, J.C. et Heethaar, R.M. (2005), \textit{The spatiotemporal MEG covariance matrix modeled as a sum
of Kronecker products}, NeuroImage, 27, 402-415.

\noindent [4] Friston, K.J., Henson, R.N.A., Phillips, C. et Mattout, J. (2005), \textit{Bayesian estimation of evoked and induced responses}, Human Brain Mapping, 27, 722-735.

\noindent [5] Mahanta, M.S.,  Aghaei, A.S. et  Plataniotis, K.S. (2012), {\it A Bayes
optimal matrix-variate LDA for extraction of spatio-spectral
features from EEG signals}, EMBC.

\noindent [6] Spinnato, J., Roubaud M.C., Burle, B. et Torrésani, B. (2014), {\it Finding EEG Space-Time-Scale localized features using matrix-based penalized discriminant analysis}, ICASSP.

\noindent [7] Farwell, L. A., Donchin, E. (1988). \textit{Talking off the top of your head: toward a mental prosthesis utilizing event-related brain potentials}, Electroen. Clin. Neuro., 70(6), 510-523.

\noindent [8] Dutilleul, P. (1999), \textit{The MLE algorithm for the matrix normal distribution}, J. Stat. Comput. Sim., 64(2),105-123. 

\noindent [9] Perrin, M., Maby, E., Bouet, R., Bertrand O. et Mattout, J. (2011), \textit{Detecting and interpreting responses to feedback in BCI}, Graz BCI International Workshop, 116-119.

\end{document}